\documentstyle[pre,aps,multicol]{revtex}

\tighten
\input epsf

\begin{document}
\draft

\title{\Large \bf  
Dynamic Approach to the \\
Fully Frustrated XY Model\thanks{
Work supported in part by the Deutsche Forschungsgemeinschaft;
DFG~Schu 95/9-1 and SFB~418}}

\author{\bf H.J. Luo, L. Sch\"ulke and B. Zheng$^*$}

\address{Universit\"at -- GH Siegen, D -- 57068 Siegen, 
Germany}

\address{$^*$Universit\"at Halle, D -- 06099 Halle, 
Germany}

\maketitle

\begin{abstract}
 Using Monte Carlo simulations, we systematically investigate
 the non-equilibrium dynamics
  of the chiral degree of freedom in the two-dimensional 
  fully frustrated XY model. The critical
initial increase of the staggered chiral magnetization is observed. 
By means of the short-time dynamics approach, 
we estimate the second order phase transition
temperature $T_{c}$ and all the dynamic and static critical
exponents $\theta$, $z$, $\beta$ and $\nu$.
\end{abstract}

\pacs{ PACS: 64.60.Cn, 75.10.Hk, 64.60.Ht, 75.40.Mg}

\begin{multicols}{2} \narrowtext
 
For critical phenomena, it is traditionally  believed that universal
behaviour exists only in equilibrium or in the long-time regime of the 
dynamical evolution. The universal scaling behaviour is described by 
a number of critical exponents. 
Due to critical slowing down, numerical measurements of 
the critical exponents are very
difficult. 

Recently, much progress has been achieved in dynamic 
critical phenomena.
It was discovered that, already after a microscopic time scale $t_{mic}$,
 universal scaling behaviour
emerges in the {\it macroscopic short-time regime}
of the dynamic process
\cite {jan89,hus89,hum91,li94,sch95,zhe98}.
To illustrate this, we consider 
the following dynamic relaxation process: a magnetic system initially in a 
high-temperature state with a small initial magnetization $m_0$
 is
quenched to the critical temperature $T_c$ 
without an external
magnetic field 
and then released to a dynamic evolution with
 model A dynamics \cite {hoh77}.
At the onset
of the evolution, the magnetization is subjected to the scaling form
\cite {jan89,li94,sch95,gra95,sch96}
\begin{equation}
M(t,\tau,m_0) \sim m_0 \, t^\theta F(t^{1/\nu z}\tau).
\label{e10}
\end{equation}
The exponent $\theta$ is a new independent exponent,
$\tau \sim (T-T_c)/T_c$ is the reduced temperature,
 $\beta$ and $\nu$ are the static critical exponents,
 and $z$ is the dynamic exponent.
At the critical temperature, $\tau=0$, the magnetization 
undergoes a {\it critical initial increase} $M(t) \sim t^\theta$. 

In fact, the short-time dynamic scaling is very general.
Another important example is the dynamic relaxation 
of a magnetic system starting
from an ordered state ($m_0=1$) \cite{sta92,sch96,sta96a,zhe98}.
The scaling form for this dynamic process is given by
\begin{equation}
M^{(k)}(t,\tau,L)=b^{-k\beta/\nu}M^{(k)}(b^{-z}t,b^{1/\nu}\tau,b^{-1}L),
\label{e20}
\end{equation}
where, for later convenience, a system with finite size $L$
has been considered. This finite size scaling has the same form as 
in the long-time regime but is assumed valid 
in the macroscopic short-time regime.

One of the most prominent properties of the scaling forms 
(\ref {e10}) and (\ref {e20}) is that 
the exponents $\beta$, $\nu$ and $z$ take the same values
as in equilibrium or in the long-time regime of the dynamic
evolution. 
It has been suggested that it is possible 
to determine not only the dynamic but also
all the {\it static} exponents as well as
the {\it critical temperature} already in the short-time regime
 \cite {sch95,sch96} (see also Refs. 
 \cite {sta96a,li96,blu92,zhe98}).
 The method may be an alternative way for overcoming 
 critical slowing down since the measurement does not
 enter the long-time regime.
It is important to systematically verify 
 this application
 in general and complex models.

The two-dimensional fully frustrated XY (FFXY) model has been the 
topic of many
recent studies
\cite{lee91,nic91,ram92,gra93,lee94,ram94,lee95,ols95,jos96}.
Critical properties of this model are rather
unconventional.  On the square lattice, the model has two kinds of phase
transitions, i.e. the Kosterlitz-Thouless phase transition (XY-like) 
and the second order phase transition (Ising-like).
Numerical simulations of
the FFXY model suffer severely from critical slowing down.
Due to the frustration, the standard cluster algorithm
does not apply to the FFXY model. 
Most of the recent work on FFXY models supports that
these two phase transitions take place at two different temperatures, 
however, their critical properties are still not 
very clear.
For example, for the
second order phase transition 
the estimated values of the exponents $\beta$ and $\nu$ 
differ in the literature \cite{jos96,ols95,lee94,gra93}
and the critical dynamics 
has not been investigated.
It is still a matter of controversy
whether the chiral degree of freedom of the FFXY model
is in the same universality class as the Ising model
\cite{ols96,jos96,ols95,ols97,jeo97}.

In this letter we present results of systematic Monte Carlo simulations
for the short-time dynamic behaviour of the second order
phase transition in the two-dimensional FFXY model. 
We confirm the short-time dynamic scaling.
For the first time, we confidently determine
the dynamic exponents $\theta$ and $z$. 
Based on the short-time dynamic scaling, the static exponents $\beta$
and $\nu$ as well as the critical temperature $T_c$ 
are also extracted from the numerical data. 

\begin{figure}[h]\centering
\epsfysize=9cm
\epsfclipoff
\fboxsep=0pt
\setlength{\unitlength}{1cm}
\begin{picture}(9,9)(0,0)
\put(0,0){{\epsffile{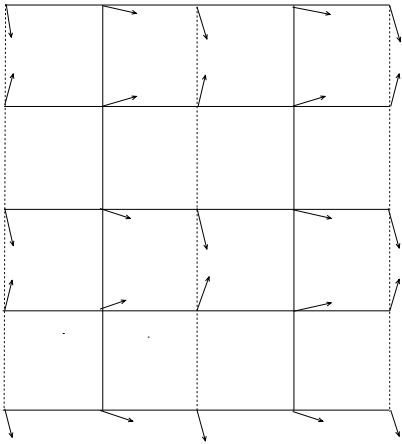}}}
\end{picture}
\caption{ One of the ground states of the FFXY model.
Dotted lines denote the links with $A_{ij}=\pi$. 
}
\label{f1}
\end{figure}

The Hamiltonian of the FFXY model on a square lattice can be written as
\begin{equation}
H=-K  \sum_{<ij>} \cos(\theta _i -\theta _j+A_{ij}),
\label{e30}
\end{equation}
In our notation the factor $1/kT$  has been absorbed in the coupling $K$,
$\theta _i$ is the angle of the spin (a unit vector) located on site
$i$, $A_{ij}$ determine the frustration and
the sum is over the nearest neighbours. 
A simple realization of the FFXY model is by taking $A_{ij}=\pi$ on  
half of the vertical links and $A_{ij}=0$ on other links.
This is shown in Fig.~\ref {f1}, where the links with $A_{ij}=\pi$
are  marked by dotted lines.  
The spin configuration of one ground state of the model
is also shown in Fig.~\ref {f1}.
 The order parameter for the second order phase transition
 is the staggered chiral magnetization defined as \cite{ram94}
\begin{equation}
M_I=\left\langle \frac {1}{L^2}  \sum_r {(-1)}^{r_x+r_y} 
   \mbox{ sgn} \left [ \sum_{
<ij> \in P_r} \sin (\theta _i-\theta_j+A_{ij}) \right] \right\rangle
\label{e40}
\end{equation}
where $(r_x,r_y)$ is the coordinate of the plaquette $P_r$.

At first, we investigate the short-time critical behaviour of 
$M_I$ in the dynamic process starting from an initial state
with a very high temperature and a small magnetization $m_0$.
We prepare an initial configuration in the following way:
first we randomly generate all spins on the lattice, then we 
randomly choose
a number of plaquettes and orient their spins
according to the configuration of the ground state shown in Fig.~\ref {f1}
till the initial magnetization $m_0$ is achieved.
After the initial configuration is generated,
the system is released to the dynamic evolution
with the Metropolis algorithm at temperatures
around $T_c$. We have performed
our simulation on lattices of size $L=128$ and $256$, 
and updated the system up to
1000 Monte Carlo steps. 
The average is taken over independent initial configurations
with $40\,000$  samples for $L=128$ and $10\,000$ samples for $L=256$.
Errors are estimated by dividing the
samples into three groups.

\begin{figure}[h]\centering
\epsfysize=8cm
\epsfclipoff
\fboxsep=0pt
\setlength{\unitlength}{1cm}
\begin{picture}(8,8)(0,0)
\put(-0.5,0){{\epsffile{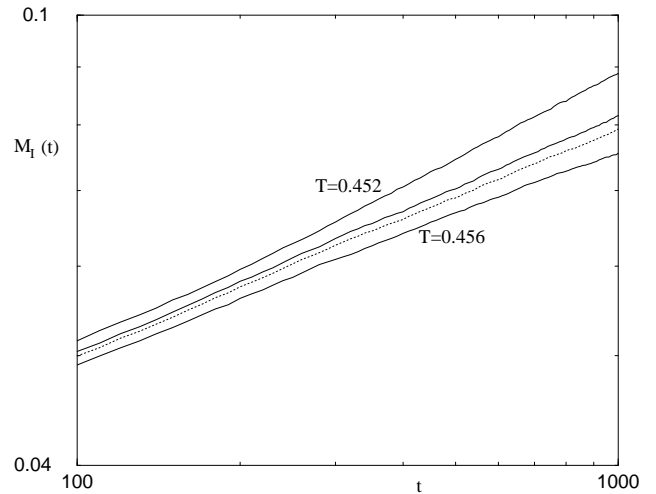}}}
\end{picture}
\caption{The time evolution of the chiral magnetization $M_I(t)$ 
when starting from a disordered initial
state. The temperatures are $T=0.452$, $0.454$, and $0.456$
from above. The dotted line represents the magnetization
at $T_c=0.4547$.
}
\label{f2}
\end{figure}

In order to locate the critical temperature $T_c$,
 simulations have been carried out
with three temperatures $T=0.452$, $0.454$ and $0.456$.
The initial magnetization is set to $m_0=0.06$.
In Fig.~\ref {f2}, the time evolution of the magnetization $M_I(t)$
at different temperatures
is plotted with solid lines in log-log scale for the lattice size $L=128$. 
Data within the microscopic time scale 
$t_{mic} \sim 100$, which are dependent on microscopic details,
are not included.
Indeed we observe that the magnetization increases
at the macroscopic early time.
The magnetization at the temperature between
$T=0.452$ and $0.456$ can be obtained by
quadratic interpolation.
From the scaling form (\ref {e10}) and 
as suggested in Ref. \cite {sch96},  searching for
a curve $M_I(t)$
with the best power law behaviour can yield an estimate of
the critical temperature. 
In Fig.~\ref {f2}, the dotted line represents such a curve 
and the corresponding critical temperature $T_c=0.4547(8)$.
From the slope of this curve,
the exponent $\theta=0.200(3)$ is obtained. 

From an analysis of the data for $L=256$ 
we have observed
that the finite size effect for $L=128$ is already sufficiently small 
to be neglected.
In fact,
the finite size effect can easily be controlled
in the short-time dynamics, and this is an
advantage of the short-time dynamic approach.
On the other hand, the critical exponent $\theta$ 
is defined in the limit $m_0 \rightarrow 0$.
Therefore, the finite $m_0$ effect should also be considered.
However, within statistical errors our data for $m_0=0.04$ and $m_0=0.06$
show no difference.
Hence the finite $m_0$ effect will also be ignored.

In principle, other 
static and dynamic critical exponents can be obtained
from $\partial _\tau \ln M_I$,
the second moment and the fourth moment
of the magnetization, the auto-correlation  and other observables.
For example,  
$1/\nu z = 0.59(3)$ is obtained from 
$\partial _\tau \ln M_I(t,\tau)|_{\tau=0}$.
However, 
this dynamic process starting from a disordered state
is not the best choice
to obtain these exponents or
the critical temperature $T_c$.
A dynamic process starting from 
an ordered state is preferable,
since the fluctuation is weaker.

For this purpose, simulations were also performed
with temperatures
$T=0.452$, $0.454$ and $0.456$, starting from
an ordered initial state.
The lattice size chosen was $L=256$ and the system 
was updated for 2000
Monte Carlo steps. The average was taken over 2000
samples. The ordered initial state was taken to be
the ground state shown in Fig.~\ref {f1}.

\end{multicols}\widetext

\begin{figure}[h]
\epsfysize=8cm
\epsfclipoff
\fboxsep=0pt
\setlength{\unitlength}{1cm}
\begin{picture}(7,7)(0,0)
\put(-1.,-0.3){{\epsffile{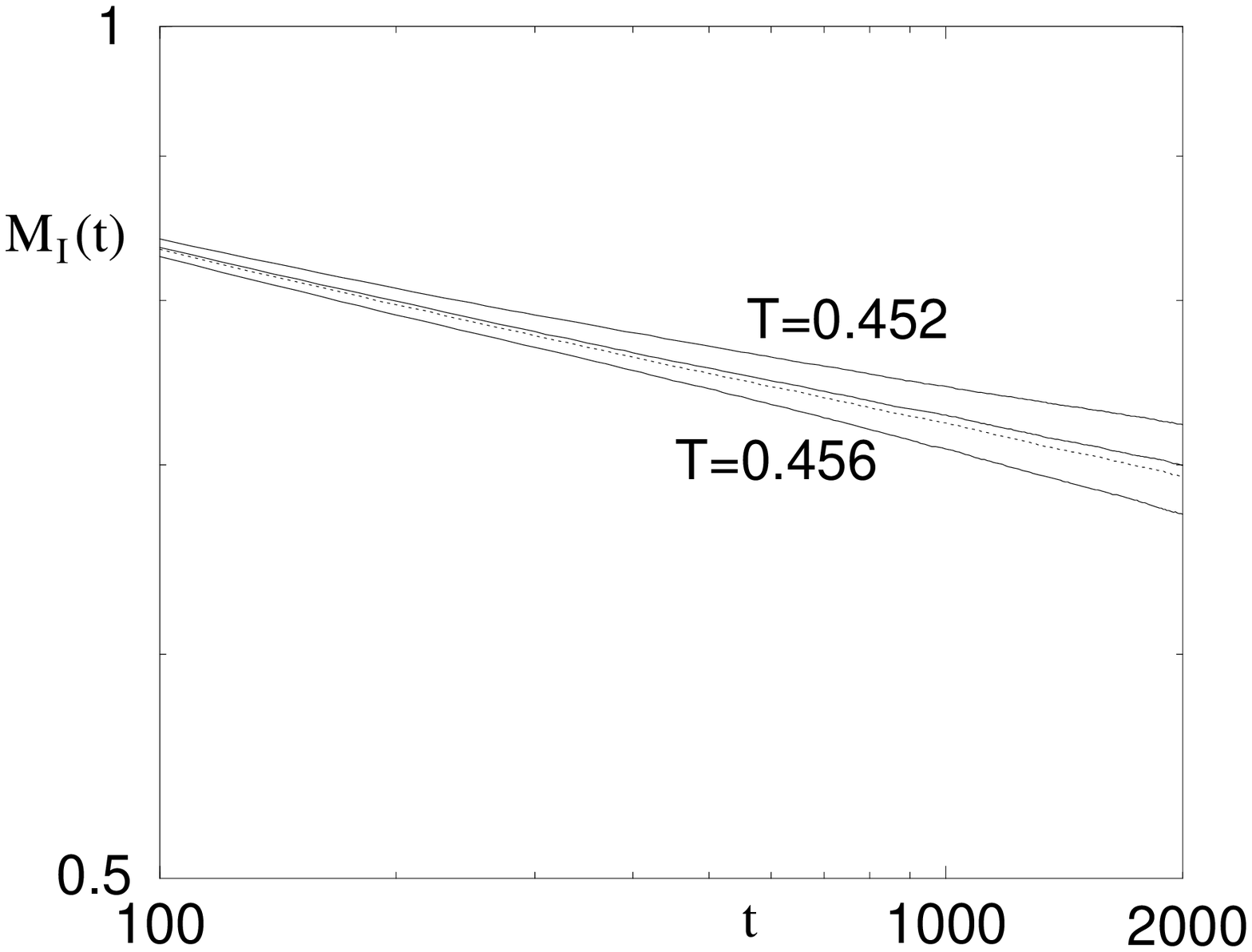}}}
\epsfysize=8cm
\put(8.5,-0.3){{\epsffile{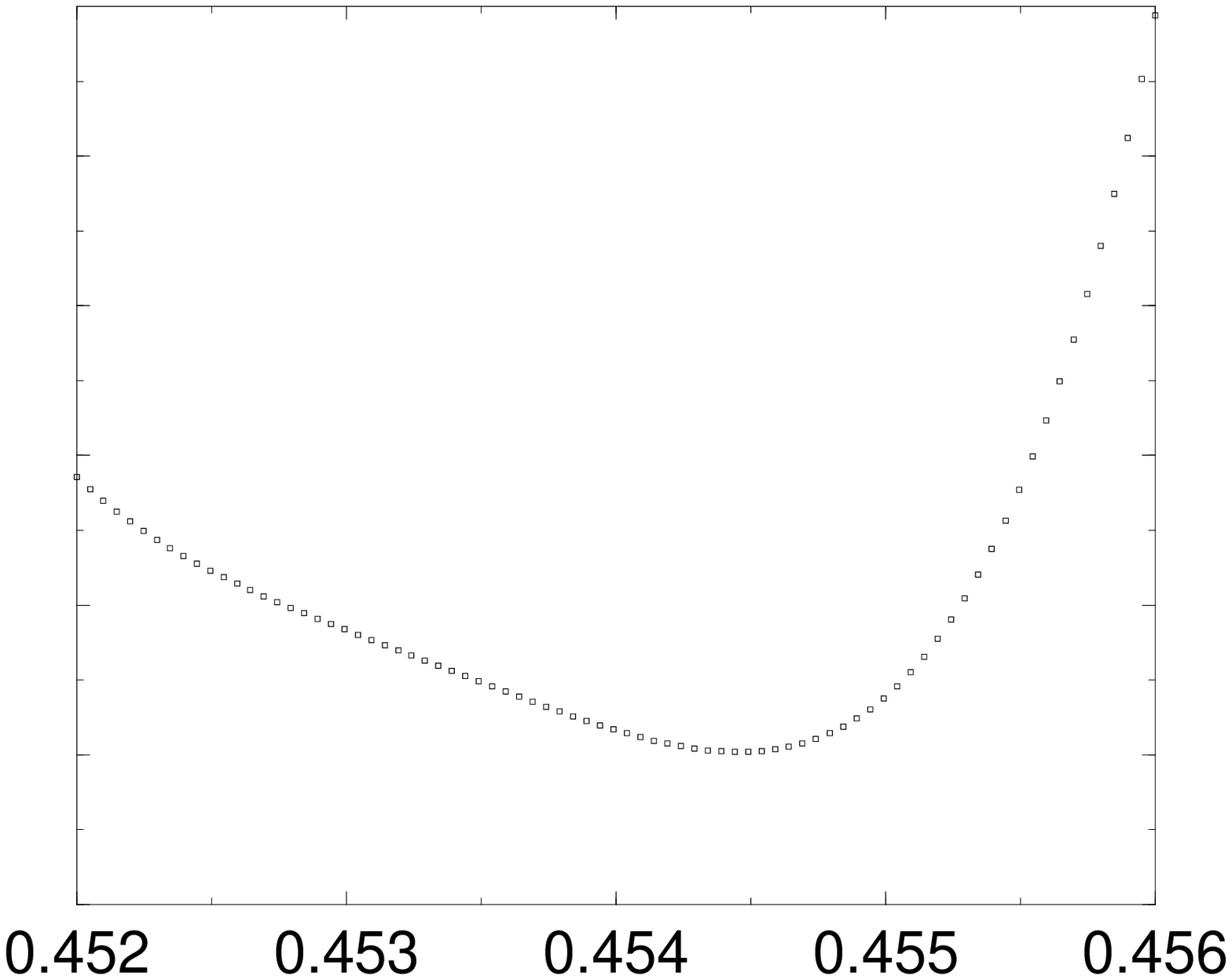}}}
\end{picture}
\caption{ (a) The time evolutions of the chiral magnetizations $M_I(t)$
when starting from an ordered state. From above, the solid lines
represent the  $M_I(t)$ at T=0.452, 0.454, and 0.456. The dotted line is
the magnetization at the critical temperature $T_c=0.4545$.
(b) Deviation of $M_I(t)$ from the power law behaviour
within the time interval $[200,1000]$. }
\label{fig3}
\end{figure}

\begin{multicols}{2}\narrowtext

The estimation of $T_c$ can now be performed again.
From the scaling form (\ref {e20}) and for 
 sufficiently large $L$
we can easily deduce the scaling behaviour for the
magnetization ($k=1$)
\begin{equation}
M_I(t,\tau)=t^{-\beta /\nu z}G(t^{1/\nu z}\tau).
\label{e50}
\end{equation}
At the critical temperature, $\tau=0$, $M_I$ undergoes a power law
decay. When $\tau \neq 0$ this power law is
modified by the scaling function $G(t^{1/\nu z}\tau)$.
As pointed out earlier, the temperature
for which the magnetization has the best power law behaviour
is the critical temperature $T_c$.
In Fig.~\ref {fig3} (a), the time evolution of $M_I(t)$ at $T=0.452$,
$0.454$ and $0.456$ is plotted in log-log scale. 
$M_I(t)$ at other temperatures
in the interval [0.452,0.456] can be estimated
by a quadratic interpolation.
To avoid the effect of $t_{mic}$, measurements
are performed in the time interval [200,2000].
The deviation from the power law can be estimated 
in different ways, e.g. as described in Ref. \cite {sch96}.
In this paper, we choose to measure the deviation 
as the error by fitting $M_I(t)$ directly
to a power law in the time interval [200,2000].
Furthermore we perform the fitting in log scale,
i.e. less weight is given to the data in the longer
time regime.
In Fig.~\ref {fig3} (b), the
deviation of $M_I$ from the power law is plotted
as a function of the temperature.
 The clear minimum confidently indicates 
the critical temperature $T_c$. 
The resulting value 
$T_c=0.4545(2)$ is consistent with $T_c=0.4547(8)$
obtained from Fig.~\ref {f2} and very close to
those values ranging from $T_c=0.451$ to $0.454$ reported in 
 most of the recent references \cite{gra93,lee94,ols95,jos96}.
Our statistical error, however, is smaller.
  The corresponding magnetization is also plotted in
Fig.~\ref {fig3} (a) with a dotted line. 
The slope of this curve yields
the critical exponent 
$\beta / \nu z = 0.0602(2)$. The quality
of this measurement is very good.
With $\beta / \nu$ given, on can obtain
a rigorous $z$ or vice versa
\cite {sta92,ito93,sta96a,sch96,zhe98}.
As compared to simulations with
a disordered initial state, 
the measurements here carry much less fluctuations.

\end{multicols}\widetext

\begin{figure}[h]
\epsfysize=7cm
\epsfclipoff
\fboxsep=0pt
\setlength{\unitlength}{1cm}
\begin{picture}(8,8)(0,0)
\put(-1.,0.3){{\epsffile{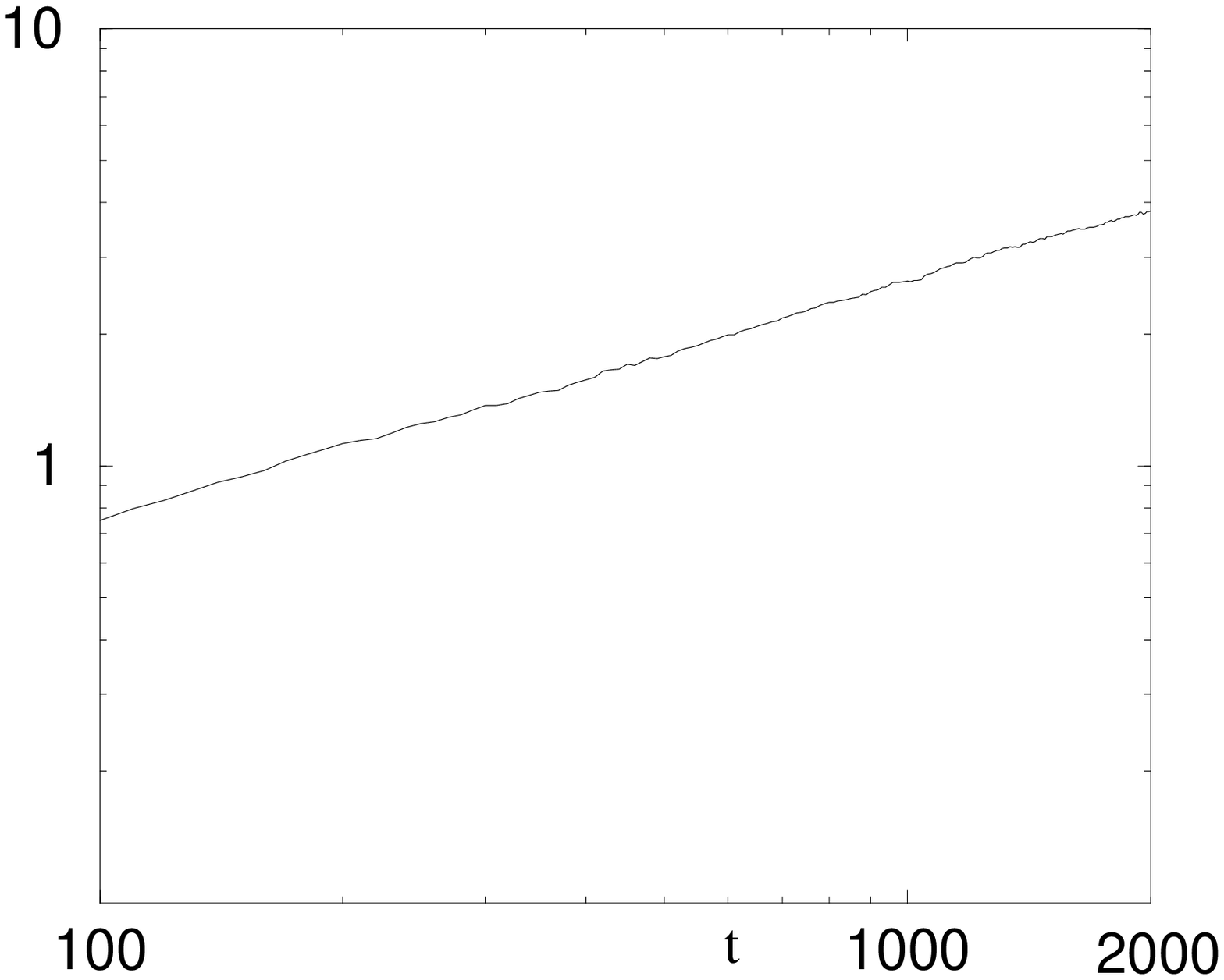}}}
\put(4.0,4){\makebox(0,0){\footnotesize $\partial_\tau \ln M_I$}}
\epsfysize=7cm
\put(9.,0.3){{\epsffile{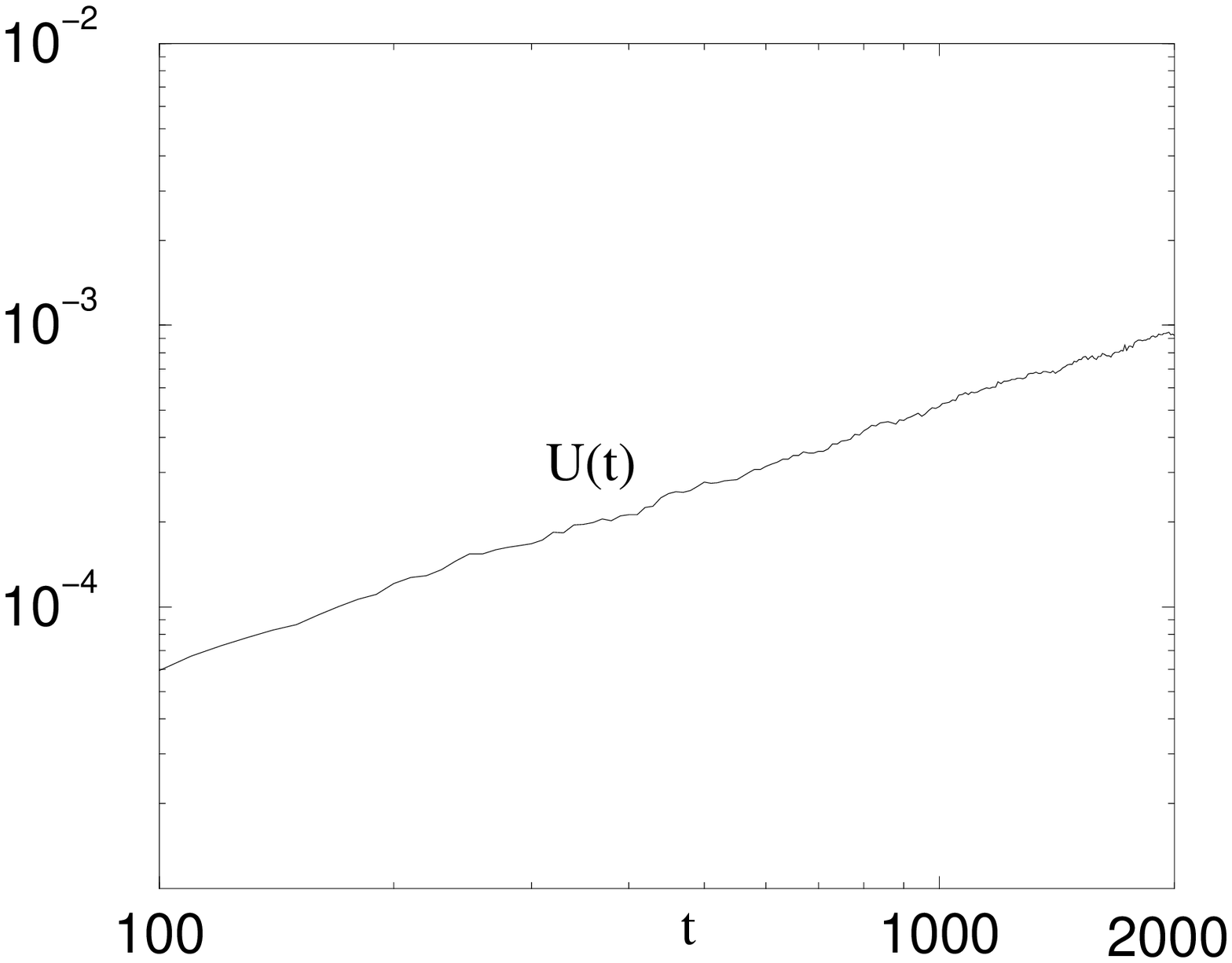}}}
\end{picture}
\caption{ (a) The derivative $\partial_\tau \ln M_I(t, \tau)|_{\tau=0}$ plotted
versus time in log-log scale.  (b) The time-dependent Binder cumulant $U$.}
\label{fig4}
\end{figure}

\begin{multicols}{2}\narrowtext

To extract the critical exponent $1/\nu z$, 
differentiation of Eq. (\ref {e50}) leads  to
\begin{equation}
\partial _\tau \ln M_I(t,\tau)|_{\tau=0}=t^{1 /\nu z}
\partial _{\tau'} \ln G(\tau')|_{\tau'=0}.
\label{e60}
\end{equation}
Therefore, $\partial _\tau \ln M_I(t,\tau)|_{\tau=0}$
should also present a power law behaviour in the beginning of the time evolution.
 In Fig. \ref {fig4} (a), 
$\partial _\tau \ln M_I(t,\tau)$ at $T_c=0.4545$ 
is plotted in log-log scale.
The power law behaviour is clearly seen.
The slope yields the critical exponent $1/\nu z=0.57(1)$.

The final step is to seek the dynamical critical exponent
$z$.  For this, we introduce a time-dependent Binder cumulant
$U(t,L)=M_I^{(2)}/M_I^2-1$. 
Due to the short spatial correlation length in the short-time regime,
a simple finite size scaling analysis shows
at the critical temperature
\begin{equation}
U(t,L)\sim t^{d/z}.
\label{e70}
\end{equation}
In Fig. \ref {fig4} (b), the curve for $U(t,L)$ in log-log scale
shows a nice power law behaviour. 
The slope gives the
critical exponent $d/z = 0.92(2)$. 
Table~\ref {t1} summarizes the results.
The errors in the measurements from $m_0=1$ are clearly smaller 
than those from $m_0=0.06$.
\end{multicols}\widetext

\begin{table}[h]\centering
\begin{tabular}{|c|c|c|c|c|c|}
       & \ $T_c$  & \ $\theta$ & \ $\beta / \nu z$ &\ $1/ \nu z$&\ $d/z$\\
\hline
 $m_0=1$   & 0.4545(2) & & 0.0602(2) & 0.57(1) & 0.92(2)\\
\hline
 $m_0=0.06$ & 0.4547(8) &  0.200(3) &  & 0.59(3) &\\
\end{tabular}
\caption{
The critical temperature and the critical exponents  
measured by simulations
starting from an ordered as well as a disordered state.
}
\label{t1}
\end{table}

\begin{table}[h]\centering
\begin{tabular}{|c|c|c|c|c|c|c|}
       & This work & Ref. \cite {jos96} & Ref. \cite {ols95}
         & Ref. \cite {lee94} & Ref. \cite {gra93} & Ising\\
       &&(1996)&(1995)&(1994)&(1993) &\\
\hline
 $T_c$   & 0.4545(2) &0.451(1) &0.452(1)& 0.454(2) & 0.454(3) &\\
\hline
 $\nu$ &0.81(2) & 0.898(3) &1  &0.813(5) &0.80(5) & 1\\
\hline
$2\beta /\nu$ &0.261(5)&&&0.22(2)&0.38(2) & 0.25 \\
\hline
$z$ &2.17(4)&&&& & 2.165(10)\\ 
\hline
$\theta$ &0.202(3)&&&& & 0.191(3)\\ 
\end{tabular}
\caption{
Critical exponents  obtained in this work and values reported
in some recent references.
Reference \protect\cite {ols95}  does not provide an estimate of the 
error on $\nu=1$.
For the Ising model, 
 exponents $\nu$ and $2\beta/\nu$ are
exact values and $\theta$ is taken from 
Refs. \protect\cite {gra95,oka97a}. The exponent $z$ 
in the literature ranges from $2.155$ to $2.172$ 
\protect\cite {zhe98,ito93,gra95,oka97a}.
Here an `average' value is given.}
\label{t2}
\end{table}

\begin{multicols}{2}\narrowtext
The interesting and important property of
the Binder cumulant is that one can estimate
independently the dynamic exponent $z$. 
With $z$ in hand, 
 we calculate the critical exponents $ 2\beta/\nu$ and
$\nu$ from $\beta/\nu z$ and $1/\nu z$.
 Table~\ref {t2} lists all critical exponents 
along
with the results reported in the recent literature.
Now $\theta=0.202(3)$ is measured at $T_c=0.4545$,
which shows a small difference from that at $T=0.4547$. 
Our short-time dynamic measurements support 
those from Ref. \cite {lee94}
and provide extra new results for 
the dynamic exponents $z$ and $\theta$.
The exponent $\nu$ of the FFXY model
is different from that of the Ising model
by nearly $20$ per cent. This indicates that
the chiral degree of freedom of the FFXY model is
in a new universality class.
Other exponents of the FFXY model do not differ much from
those of the Ising model.
From our numerical data, we observe that
the values of the critical exponents are rather sensitive to
the assumed or measured critical temperature $T_c$.
This should be one of the main reasons for the
different values of the exponents reported in
the literature. 
Figure~\ref {f2} and, in particular, Fig. \ref {fig3},
give us confidence in our measurements of the critical temperature
$T_c$. Furthermore, our exponent $\nu$ is extracted from the data
in the close neighbourhood of $T_c$, in contrast to
many simulations in equilibrium.
The finite size effect is also well under control
in the short-time dynamic approach.

In conclusions, using Monte Carlo methods,
we have systematically investigated 
and confirmed the universal short-time 
dynamic behaviour of the second order phase transition 
in the two-dimensional 
fully frustrated XY model.
Based on the short-time dynamic scaling form,
all static and dynamical critical exponents are determined. 
The dynamic exponents $\theta$ and $z$ are obtained   
for the first  time
 and the measurement of the
 exponent $\beta/\nu z$ and the critical temperature
$T_c$ is very precise.
The estimated value $\nu = 0.81(2)$ is clearly
different from $\nu=1$ for the Ising model.
Our investigation of the chiral degree of freedom
of the FFXY model is to date the most systematic.
 We are convinced that the short-time
dynamic approach is not only conceptually interesting 
 but
also practically efficient.
In the simulations we do not encounter difficulties
associated 
with large correlation times since 
our measurements are carried out
in the short-time regime and we do not have the problem of
generating
independent configurations.

A possible extension of the present work is 
the Kosterlitz-Thouless phase transition.
In fact, some exponents have been obtained
\cite {luo97a,luo97}. However, owing to the absence of
symmetry breaking, a
clear signal such as
in Fig. \ref {fig3} does not exist
 for the Kosterlitz-Thouless transition temperature $T_{KT}$.
The determination of $T_{KT}$ and of the exponent $\nu$
is very difficult and
requires extensive simulations and a careful analysis.


\end{multicols}\widetext

\end{document}